# Noise performance and long-term stability of near- and mid-IR gas-filled fiber Raman lasers

Yazhou Wang,[1,*] Abubakar I. Adamu,[1] Manoj K. Dasa,[1] J. E. Antonio-Lopez,[2] Md. Selim Habib,[2,3] Rodrigo Amezcua-Correa,[2] Ole Bang,[1,4,5] and Christos Markos[1,4,*]

[1]*DTU Fotonik, Technical University of Denmark, DK-2800 Kgs. Lyngby, Denmark*
[2]*CREOL, The College of Optics and Photonics, University of Central Florida, Orlando, FL-32816, USA*
[3]*Department of Electrical and Computer Engineering, Florida Polytechnic University, Lakeland, FL-33805, USA*
[4]*NORBLIS IVS, Virumgade 35D, DK-2830 Virum, Denmark.*
[5]*NKT Photonics A/S, Blokken 84, Birkerød, DK-3460, Denmark*
*\*Corresponding authors: yazwang@fotonik.dtu.dk;chmar@fotonik.dtu.dk*

**In this letter, the characteristics of noise and long-term stability of near- and mid-infrared (near-IR and mid-IR) gas-filled fiber Raman lasers have been investigated for the first time. The results reveal that an increase in Raman pulse energy is associated with a decrease in noise, and that the relative pulse peak intensity noise (RIN) is always lower than the relative pulse energy noise (REN). We also demonstrate that long-term drift of the pulse energy and peak power are directly linked with the high amount of heat release during the Raman Stokes generation. The demonstrated noise and long-term stability performance provide necessary references for potential spectroscopic applications as well as further improvements of the emerging mid-IR gas-filled hollow-core fiber (HCF) Raman laser technology.** ©2020 Optical Society of America

https://doi.org/10.1364/OL.99.099999

The emerging gas-filled HCF Raman laser technology has attracted enormous attention due to its compelling features that enable efficient stimulated Raman scattering (SRS) process, high damage threshold, broad transmission range, compact fiber structure, high freedom on wavelength selection, etc. [1-3]. Especially, with the recent advent of low loss anti-resonant HCF technology, the laser wavelength has been extended to the mid-IR region where the high silica loss could be significantly mitigated by confining most power of the laser beam within the fiber core (gas) region [4, 5]. A series of gas-filled fiber Raman lasers have been reported from UV to mid-IR region. Despite their unique advantages compared to other counterparts, high noise is an intrinsic issue of Raman lasers since the SRS process is usually initiated by the quantum noise. The noise of Raman laser has extensively been investigated in the frame of bulk gas cell usually filled by hydrogen ($H_2$) [6], but yet remains unexplored within the gas-filled fiber technology. The noise characteristics of the later is worth further investigation since the SRS process in gas-filled HCF is much more efficient than the conventional bulk gas cell structure. Besides, for the Raman Stokes emission involving a large photon energy difference between the pump and Stokes, the energy of the released phonon is high and thus could easily lead to a large amount of heat release within the core region of the HCF, where the gas circulation speed is much slower than its counterpart of using bulk gas cell. As a result, the temperature and therefore the Raman gain coefficient will be directly affected. In the last few years, several efficient Raman lasers operating at ~ 4 μm wavelength have been reported with $H_2$-filled anti-resonant HCFs (ARHCFs) by using 1.5 μm fiber laser as a pump [4, 7, 8], where the generation of one Stokes photon is accompanied by one phonon emission with up to ~60 % of pump photon energy. Consequently, the heat pulse energy of these mid-IR Raman lasers is calculated to be in the microjoule level, which could impose a significant impact on the laser stability. However, as an emerging laser technology, these issues have not been investigated yet.

In this letter, we thoroughly investigate these noise effects based on our previous work [8], where a 4.22 μm Raman line with a high quantum efficiency of ~ 70 %, pulse energy of ~ 16 μJ, and few nanosecond pulse duration was reported based on a $H_2$-filled ARHCF with 15 bar pressure, pumped by a custom-made linearly polarized 1532.8 nm fiber laser with a pulse energy of ~ 80 μJ and pulse duration of 7 ns. Figure 1 shows the details of the experimental setup used for this investigation. The output from the $H_2$-filled ARHCF is first collimated and then the residual pump is removed by a long-pass filter (LPF). The remaining 4.22 μm Raman part is passed through a $CaF_2$ window (~ 95 % transmission at 4.22 μm) and monitored by a thermal power meter. A small part of the Raman laser reflected from the front surface of the $CaF_2$ window is focused onto a photodetector (PDAVJ10, Thorlabs) which is connected to an oscilloscope (Teledyne Lecroy HD09000) for pulse profile monitoring. The pulse duration of the Raman laser is estimated to be a few nanoseconds, which is comparable with the rise time of the photodetector (~ 3.5 ns), as a result, the Raman laser pulse profile couldn't be accurately measured. Here we use the deconvolution method to retrieve the original pulse profile, as described in Supplement 1. Because the 4.22 μm Raman line here has a high polarization extinction ratio (PER) (> 20 dB) (see Supplement 2), a polarizer was used to continuously attenuate the average power, so that most of pulses could be appropriately measured in the linear response region of the photodetector. The characterization of the photodetector is provided in Supplement 1. The pump laser was also simultaneously monitored by focusing a minor part of the pump into a near-IR photodetector (DET01CFC, Thorlabs).

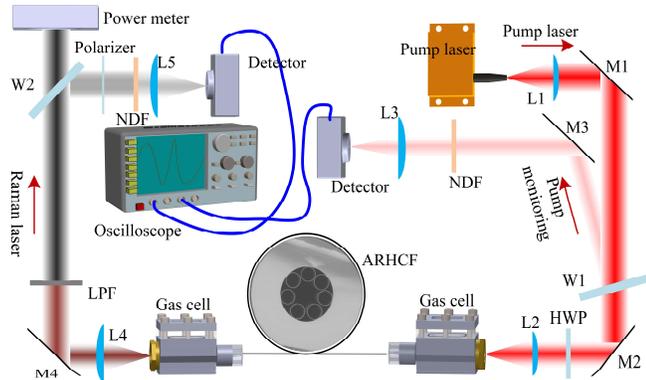

Figure 1. Noise measurement setup of a mid-IR gas-filled fiber Raman laser. M1-M4 are silver mirrors. L1-L3 are C-coated plano-convex silica lenses, L4-L5 are uncoated plano-convex $CaF_2$ lenses. W1 and W2 are respectively optical wedge and window used to extract a small part of the laser beam for pulse detection with photodetectors.

From the setup in Fig. 1, due to the fact that the pump is linearly polarized, the rotational Raman line at 1.68 μm has a near-zero power (see the measured spectrum in Fig. 2) and thus negligible influence on the generation of an vibrational Raman line at 4.22 μm. Nevertheless, for the purpose of comparison, we also monitored the noise and long-term stability of the rotational Raman line generated at 1.68 μm when replacing the half-wave plate (HWP) with a quarter-wave plate (QWP) in front of the ARHCF to make the pump light circularly polarized [9]. When using circularly polarized pump light, the $H_2$ pressure was reduced to 8 bar to suppress the vibrational Raman line, which finally provided a quantum efficiency and pulse energy of the rotational Raman line of 45 % and 27 μJ, respectively, while the average power of the vibrational Raman line was measured to be less than 1 mW (see Fig. 2). Then, the pulse profile of the rotational Raman line was precisely measured by the other near-IR photodetector with the same parameters as the one used for monitoring the pump. Because of the circular polarization of the rotational Raman line, the polarizer was replaced by a continuously variable near-IR neutral density filter (NDF) (NDL-10C-2, Thorlabs), to precisely attenuate the pulse peak power to the linear region of the photodetector.

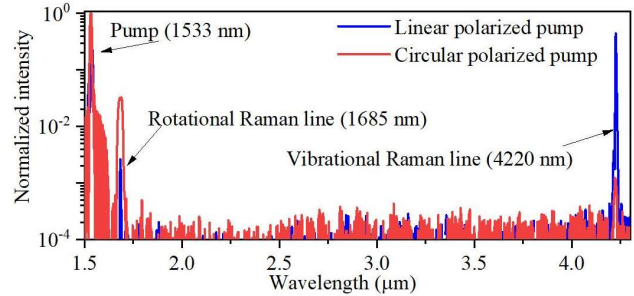

Figure 2. Spectra comparison of the Raman lines with linear and circular polarized pumps.

In the noise measurement, we collected a train of 10,000 Raman pulses and corresponding pump pulses, for each pump power level. The influence of the pump noise is suppressed by discarding the pump pulses whose peak intensity fall outside 2 % of its average value. After the data processing, the number of pairs is reduced to ~9500, and the relative noise (defined as the ratio of the standard deviation to the mean value) of the pump peak intensity is reduced from ~1.7 % to ~1 %. Figure 3(a) and 3(b) present the measured REN and RIN of the vibrational and rotational Raman lasers. It shows that both of them gradually decrease and finally approach a stable level as the Raman laser power increases. This is attributed to the increase in the efficiency of the SRS process and the Raman gain saturation [10]. Meanwhile, at the same pulse energy level, it can be seen that the RIN is always lower than the REN. For instance, at the highest power of 132.8 mW in Fig 3(a), the REN is 16.8 %, while the RIN is as low as 2.9 %. This difference is supposed to be caused by the dependence of the SRS efficiency on the light intensity: the SRS process in the peak region of the Raman pulse is more efficient than at its pulse edge regions, while high SRS efficiency is associated with low noise level. More details about this deduction are provided in Supplement 3.

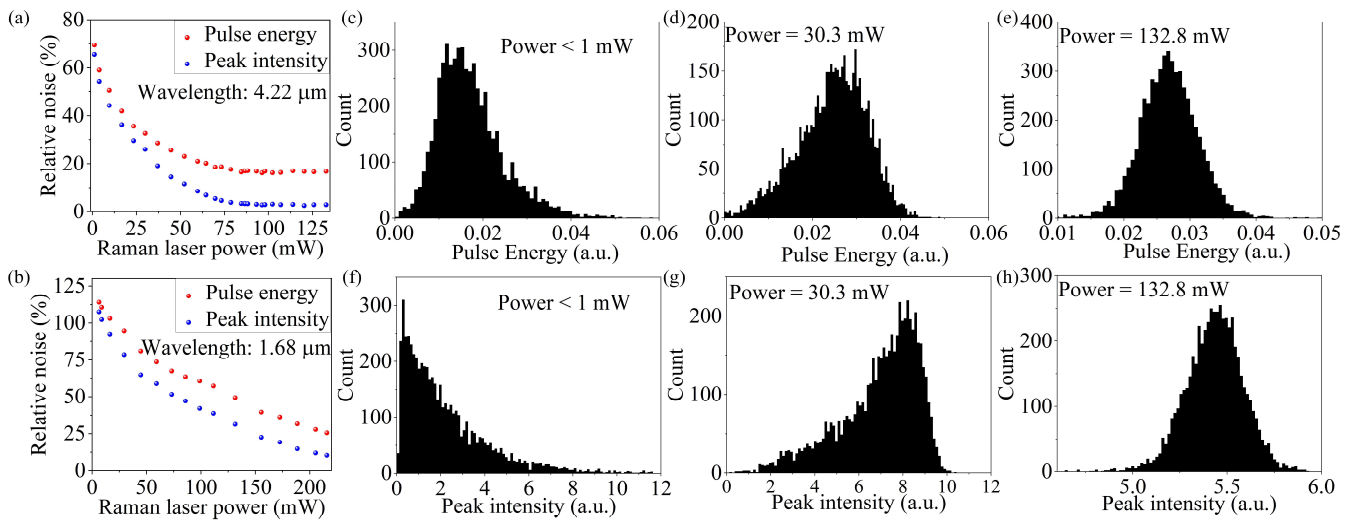

Figure 3. (a) and (b) are the measured REN and RIN of the 4.22 μm and 1.68 μm Raman lines as a function of their average power, respectively. (c)-(e) are histograms of measured pulse energy at different levels of output Raman laser power (4.22 μm). (f)-(g) are histograms of measured pulse peak intensity corresponding to (c)-(e).

Figure 3(c) shows a histogram of the pulse energy of the vibrational Raman line at low Raman laser power (of less than 1 mW), where the distribution approaches a negative exponential distribution. If one compares with the reported experimental results in [6, 10], this distribution slightly deviates from an ideal exponential distribution, which might be attributed to the influence of the relatively high background noise of the photodetector. A shred of evidence for this deduction is the comparison of the corresponding histogram of the peak intensity in Fig. 3(f), where the distribution is even closer to an exponential distribution due to the less influence of the background noise on the peak intensity when compared to the pulse area. When the laser power increases, the distributions of both pulse energy and peak intensity gradually evolve towards a symmetrical Gaussian-like distribution with a narrower width, as shown in Fig. 3(c)-3(e) as well as Fig. 3(f)-3(h). This evolution is consistent with the theoretical prediction and other experimental results of the conventional Raman lasers based on a bulk gas cell configuration [6, 10].

As mentioned in the introduction, the generation of one 4.22 µm Stokes photon from its 1.53 µm pump is accompanied by high energy phonon release. Here, the average heat release from each pump pulse at the quantum efficiency of ~70 % (at the pressure of 15 bar) is as high as ~25 µJ. In terms of spatial distribution, heat energy is supposed to be mainly concentrated in the end part of the ARHCF due to the rapid increase of Stokes pulse energy in this region [7]. Therefore, the temperature field is supposed to have a non-uniform distribution along the fiber, consequently leading to a gas circulation inside the laser system. Meanwhile, since Raman lines initiated by quantum noise have a relatively high fluctuation, the heat release from each pump pulse consequently fluctuates also. The combination of gas circulation and fluctuation of the released heat energy could easily result in an irregular variation of the gas-filled ARHCF temperature and thus the Raman gain coefficient, thereby compromising the long-term stability of the laser delivering. In order to verify this, we first observed the long-term stability at two different power levels, as shown in Fig. 4(a), where the fluctuation of the pulse peak intensity was monitored over ~ 30 minutes by recording 120,000 pulses with 144 ms average time separation between two adjacent pulses. The long-term stability measurement was implemented after 1 hour of warming-up the laser system, to ensure that the system had reached a stable state. It can be seen that the peak intensity of the Raman laser at the power level of 132.8 mW exhibits obvious drift (red curve in the bottom, right) when compared to that observed when using less than 1 mW power (grey curve in the bottom, left), while the pump laser fluctuation remains at the similar level without obvious drift (curves in the top). Note that a similar drift could be observed in terms of pulse energy, and these phenomena are highly repeatable in our experiment. The Allan deviation, which is a numerical model widely used for the evaluation of a system's stability [11, 12], is introduced here to quantitatively describe the drift effect, as shown in Fig. 4(b), where the Allan deviation of the pulse peak intensities are presented at four power levels. Initially, because it is white noise dominated, the Allan deviation linearly decreases with increasing time [11]. When the time is sufficiently long to include the drift effect, the Allan deviation gradually deviates from the linear decrease and finally starts to increase again, i.e., entering the drift-dominated region [11]. The time $t_{min}$ at the point of inflection is therefore a sign of the magnitude of drift. Here $t_{min}$ is ~106 s at a Raman laser power of less than 1 mW and it then gradually decreases to 48 s when the power increases to 132.8 mW, indicating that obvious drift is induced when increasing the Raman laser power.

Note that the pump drift maintains a similar level at different powers. Figure 4(c) presents the Allan deviations of the pump laser corresponding to Fig. 4(b). It can be seen that, although their values at the initial time of $t$ = 14.4 ms have a slight difference, they evolve with similar trend and gradually overlap with each other. In order to exclude the influence of the pump drift on the drift of the Raman laser, we discarded the pulse pairs where the pump peak intensity falls outside 0.1 % of its average value. Then, the time separation of adjacent measured pulses is still set to 14.4 ms, to calculate the Allan deviation. The pump Allan deviations after the data processing are presented in Fig. 4(d), where their powers are corresponded with the Raman laser powers of 132.8 mW and < 1 mW, respectively. Both Allan deviations linearly decrease as a function of time, indicating that the drift is effectively mitigated after the data processing. In this case, as shown by the red and green curves in Fig. 4(d), the Allan deviation of Raman laser with 132.8 mW power still exhibits obvious inflection point at $t_{min}$ = 43 ms, while linearly decreases at the power less than 1 mW as a result of the low amount of heat release. This suggests that the high heat release caused by the high Raman laser energy indeed induces the drift of the Raman laser.

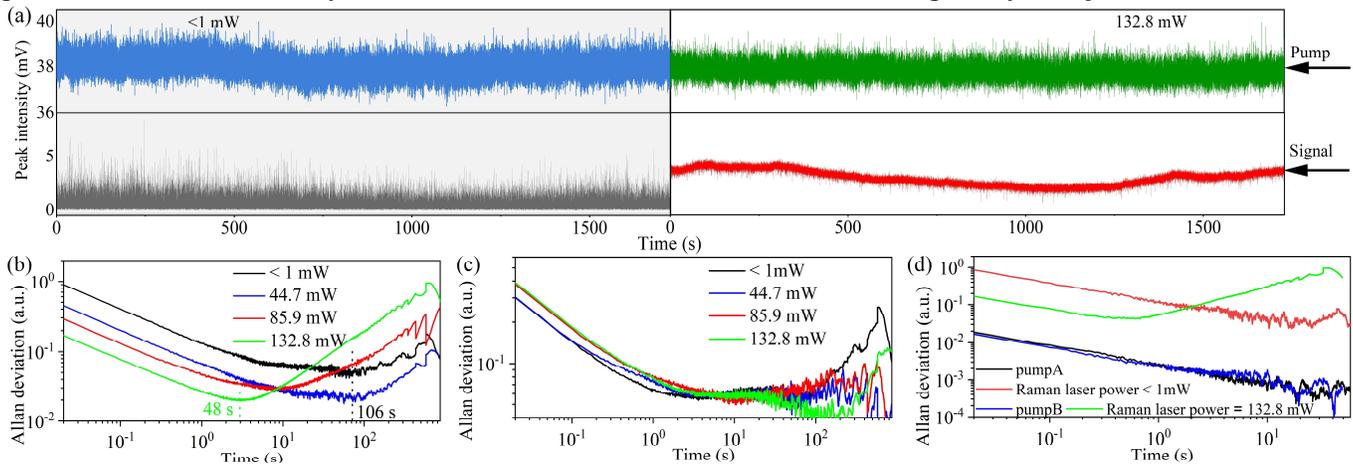

Figure 4. (a) Long term stability monitoring of the pulse peak intensity of the pump laser (top) and corresponding vibrational Raman laser (bottom) at two different Raman laser powers of < 1 mW (left) and 132.8 mW (right). (b) Allan deviation of the vibrational Raman pulse peak intensity at different Raman laser power levels. (c) Allan deviation of the pump pulse peak intensity corresponding to (b). (d) Allan deviations of pump pulse intensity by filtering out their drift parts, as well as the corresponding Allan deviations of the Raman lasers (PumpA and pumpB in (d) donate the pump corresponding to the Raman laser with powers of less than 1 mW and equal to 132.8 mW, respectively).

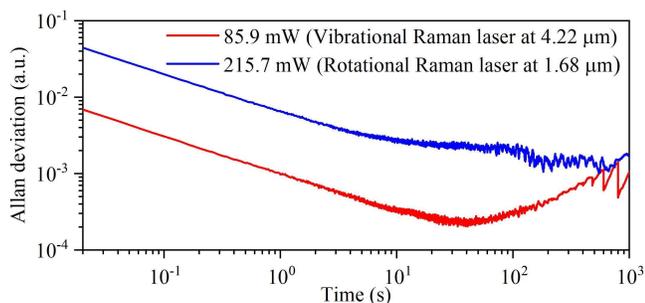

Figure 5. Comparison of the Allan deviations of the vibrational and rotational Raman lasers under the same quantum efficiency of ~45 %.

The other experimental evidence for the heat-induced drift effect is the comparison of the Allan deviations between the rotational and vibrational Raman lasers, where the Stokes photon generation at 1.68 μm is associated with low heat release and only accounts for ~9 % pump photon energy. Figure 5 shows a comparison of the two Allan deviations at the same quantum efficiency of ~45 %, i.e., 85.9 mW for the vibrational Raman laser and 215.7 mW for the rotational Raman laser. Different from the Allan deviation of vibrational Raman laser, the rotational Raman laser has its Allan deviation continuously decreasing as a function of time. This implies that the rotational Raman laser here has a weak drift due to its small amount of heat release when compared to the vibrational Raman laser.

The noise and drift of the Raman laser could compromise its potential applications. For instance, in the field of gas sensing and monitoring, a long integration time is required to offset the influence of the strong noise of the Raman laser, which however slows down the speed of gas detection [11, 13]. On the other hand, the laser drift inevitably induces a drift of the detection signal in gas detection, which in turn influences the detection limit of the target gas [11, 14]. The Raman laser drift induced by the heat release could be mitigated by avoiding the generation of the phonon, which can be achieved through satisfying the phase matching condition of the generation of anti-Stokes photon [15]. However, it is usually difficult to satisfy the conditions of both phase matching and the efficient emission of mid-IR Raman Stokes with the same fiber parameters and gas pressure. The other possible way to mitigate the Raman laser noise is to suppress the laser fluctuation by reducing the effect of quantum noise by using an external seed. Here, the noise of the 4.22 μm vibrational Raman laser could be effectively suppressed by seeding it with a mid-IR quantum cascade laser which is known for its low-noise level and narrow linewidth.

In conclusion, we investigated the noise and long-term stability of an $H_2$-filled Raman laser in both the near- and mid-IR spectral regions. Both REN and RIN of this Raman laser continuously decrease with increasing pulse energy. At the same Raman laser power level, the RIN is always lower than the REN. The mid-IR Raman laser emission is associated with a large amount of heat release and long-term stability monitoring demonstrated some drift in the pulse peak intensity and pulse energy over time. Both the relatively high noise level and long-term drift could be underlying issues for the emerging mid-IR gas-filled Raman lasers, despite their promising advantages of high quantum efficiency and both high energy and high power laser emission.

**Funding.** This work is supported by the Danmarks Frie Forskningsfond Hi-SPEC project (Grant No. 8022-00091B), Innovation Fund Denmark UVSUPER (Grant No. 8090-00060A), Innovation Fund Denmark ECOMETA (Grant No. 6150-00030B) and US ARO (Grant No. W911NF-19-1-0426).

**Acknowledgment.** The authors would like to thank Callum R. Smith and Shreesha Rao Delanthabettu Shivarama for discussions of the noise measurement. The authors also thank Christian R. Petersen for providing the mid-IR OPA.

**Disclosures**. The authors declare no conflicts of interest.

**Supplementary Materials.** See Supplement 1-3 for supporting content.